\documentclass{sig-alternate-05-2015}

\usepackage{booktabs}
\usepackage{subcaption}

\usepackage{dsfont}

\usepackage{float}

\usepackage{booktabs}

\begin{document}

\title{Routing Memento Requests Using Binary Classifiers}
%
%
%
%
%

\numberofauthors{3} 
%
\author{
%
%
Nicolas J. Bornand\\
       \affaddr{Los Alamos National Lab}\\
	\affaddr{Los Alamos, NM, USA}\\
       \email{nbornand@lanl.gov}
\alignauthor
Lyudmila Balakireva\\
       \affaddr{Los Alamos National Lab}\\
       \affaddr{Los Alamos, NM, USA}\\
       \email{ludab@lanl.gov}
\alignauthor
Herbert Van de Sompel\\
	\affaddr{Los Alamos National Lab}\\
	\affaddr{Los Alamos, NM, USA}\\
	\email{herbertv@lanl.gov}
}

\maketitle
\begin{abstract}
The Memento protocol provides a uniform approach to query individual web archives. 
Soon after its emergence, Memento Aggregator infrastructure was introduced that 
supports querying across multiple archives simultaneously. 
An Aggregator generates a response by issuing the respective Memento request against each of the distributed archives it covers. 
As the number of archives grows, it becomes increasingly challenging to deliver aggregate responses while keeping 
response times and computational costs under control. Ad-hoc heuristic approaches have been introduced to address this challenge and 
research has been conducted aimed at optimizing query routing based on archive profiles. In this paper, we explore the use of binary, archive-specific classifiers 
generated on the basis of the content cached by an Aggregator, to determine whether or not to query an archive for a given URI. 
Our results turn out to be readily applicable and can help to significantly decrease both the number 
of requests and the overall response times without compromising on recall. 
We find, among others, that classifiers can reduce the average number of requests by 77\% compared 
to a brute force approach on all archives, 
and the overall response time by 42\%  while maintaining a recall of 0.847.
\end{abstract}



\section{Introduction}
The Memento ``Time Travel for the Web'' protocol was first introduced in 2009 \cite{mementofirst} and its formal specification 
was concluded in December 2013 with the publication of RFC7089 \cite{rfc7089}. The protocol specifies interoperable 
access to resource versions, named Mementos, and consists of two complimentary components:
\begin{itemize}
\item A TimeGate (URI-G) associated with an Original Resource (URI-R) supports datetime negotiation - a variant on content negotiation - 
to allow access to a Memento (URI-M) for the Original Resource that was the live web version at or around a preferred datetime. That datetime is expressed 
in a special-purpose HTTP protocol request header.
\item A TimeMap (URI-T) associated with an Original Resource (URI-R) provides an overview of all Mementos for an Original Resource 
known to the system that provides the TimeMap. For each such Memento, the TimeMap lists the URI-M and the archival datetime.
\end{itemize}

The Memento protocol can be adopted by web archives and resource versioning systems. At the time of writing, 
especially the former systems support the protocol either through native or by-proxy implementations. 
As such, it has become possible to uniformly interact with web archives 
in order to determine which Mementos a specific archive holds for a given URI-R (TimeMap component) as well as  
to negotiate access to the Memento for a given URI-R that is held by a specific archive and that 
is temporally closest to a preferred datetime (TimeGate component). In addition, 
in order to provide these same functionalities across archives, Memento Aggregator infrastructure has been 
introduced that provides TimeMaps and TimeGates that cover multiple archives. 

\begin{table}[t]
        \caption{Web archives covered by the LANL Aggregator}
        \label{table:ArchiveCoverage}

    \centering
    \begin{tabular}{cccc}
        \toprule
        Abbreviation  & Native - By Proxy & Included \\
        \midrule
            archive.is & native & yes \\                                                  
            archiveit & native & yes \\
            ba & native & yes \\
            blarchive & native & yes \\
            es & by proxy & yes \\
            gcwa & by proxy & yes \\
            hr & by proxy & yes \\
            ia & native & yes \\
            is & native & yes \\
            loc & native & yes \\
            nara & by proxy & no \\
            proni & native & yes \\
            pt & by proxy & yes \\
            sg & by proxy & yes \\
            si & by proxy & no \\
            swa & native & yes \\
            uknationalarchives & native & yes \\
            ukparliament & native & yes \\
            webcite & by proxy & yes \\
        \bottomrule
    \end{tabular}
\end{table}

The longest running Memento Aggregator infrastructure is operated by the Research Library at the Los Alamos National Laboratory (LANL). 
As shown in Table \ref{table:ArchiveCoverage}, it currently covers 19 archives\footnote{Full archive names at \url{http://mementoweb.org/depot/}}, 
11 of which are natively Memento compliant, 
and the 8 others are compliant via proxy implementations. 
The last column in the Table indicates whether an archive was included in the experiments described in this paper. 
This Aggregator infrastructure is leveraged to deliver end user web time travel services 
(e.g. Memento for Chrome\footnote{\url{http://bit.ly/memento-for-chrome}}, 
the Time Travel web portal\footnote{\url{http://timetravel.mementoweb.org}},
Mink\footnote{\url{http://matkelly.com/mink}}) 
 and is also frequently used for research endeavors that require cross-archive lookups. 
 The Aggregator received about 1.5M incoming TimeGate/TimeMap requests in March 2015, nearly 18.5M in October 2015, and 
 over 50M in December 2015. 

In essence, the Aggregator infrastructure accepts 
TimeGate and TimeMap requests and provides responses that reach across all covered archives. 
Generating a response requires issuing the respective Memento request against each of the distributed archives. 
Since doing so is predictably resource intensive and time consuming, an Aggregator Cache has been introduced. 
The cache has URI-R as key and cross-archive TimeMap information (URI-Ms and associated archival datetimes) as value. 
 The URI-Rs that are covered by the cache are a combination of about 500K popular URIs retrieved from Alexa\footnote{\url{http://www.alexa.com/}} 
 in December 2014 plus URIs that were requested by users over time, for a total of about 1.2M. 
 
 On a recurrent basis, and in a background process, the cache is refreshed by re-polling all covered web archives for TimeMaps. 
 TimeGate/TimeMap requests against the Aggregator for any given URI-R are served 
 from the cache if the URI-R exists in the cache and 
 the cache is not considered stale. For responses that can not be delivered from cache (i.e. cache misses), the following approach is currently taken:
\begin{enumerate}
\item A TimeGate response is generated by issuing a realtime TimeGate request against each of the Memento compliant archives, 
excluding by-proxy compliant ones. The exclusion is aimed at reducing response times and required 
computational resources, and is informed by the intuition that 
responses from by-proxy implementations will generally be slower than those from native ones. 
Depending on the application, all TimeGate responses are returned to a client of 
the Aggregator or only the response with the Memento that has an 
archival datetime closest 
to the requested preferred datetime.
\item A TimeMap response is generated by issuing a realtime TimeMap 
request against all covered archives, both compliant and by-proxy, 
merging all responses, and returning them to a client of the Aggregator. This approach may yield 
significant response times but aligns with the Memento protocol that emphasizes completeness of TimeMap responses. \\ \\
\end{enumerate}

\section{Problem Statement}

The use of a cache for LANL's Memento Aggregator and the heuristic introduced for handling TimeGate 
requests for URI-Rs that are not cached are indicative of a general challenge related to operating Memento Aggregator 
infrastructure. As the number of web archives increases, 
delivering aggregate responses becomes more challenging 
as there is a limit to the number of archives that can be polled when response times and computational costs 
for the infrastructure are a concern. But, equally important is appropriately handling the load caused by requests on the 
individual archives. This may not be a serious concern in case of the Internet Archive that has 
sufficient machine power to handle continuously high traffic from around the globe.  
But, other archives have more limited resources and sometimes even policies aimed at reducing 
traffic. For example, in recent 
Hiberlink\footnote{\url{http://hiberlink.org}} research, 
we experienced a daily cap on the number of requests from a given IP address imposed by the webcite archive. 
And, soon after the overwhelmingly successful launch of oldweb.today\footnote{\url{http://oldweb.today}} in December 2015, several archives 
struggled with the load incurred by the service, leading to extreme response times 
and even a request from an archive not to be polled. 
For these reasons, Memento Aggregator infrastructures are in need of strategies that 
inform selective polling of archives instead of brute force 
polling of all archives. This consideration is supported by Table \ref{table:ArchiveHistogram}, 
which shows that 82.23\% of URI-Rs covered by the LANL Aggregator have Mementos in 0, 1, or 2 archives only. 
Clearly, using a brute force strategy, many request are issued that do not return Memento information. 
But how to know which URI-R to look up in which archive? 
How to predict whether an archive has Mementos for a given URI-R?

\begin{table}
    \caption[]{Distribution of the cached URI-R across archives}
    \centering
    \begin{tabular}{ccc}
        \toprule
            k & \# URI-R stored by k archives & In \% \\
        \midrule
            0 &  270,495 & 22.17  \\
            1 & 407,998 & 33.44  \\
            2 & 323,596 & 26.52  \\
            3 & 120,829 & 9.90  \\
            4 & 53,212 & 4.36 \\
            5 & 25,947 & 2.12 \\
            6 & 11,819 & 0.97 \\
            7-19 & 6,100 & 0.50 \\
        \bottomrule
        \label{table:ArchiveHistogram}
    \end{tabular}
\end{table}


Considering the limitations of prior work in this realm (see Section \ref{section:RelatedWork}), 
we set out to explore whether a machine learning approach could be used to inform the decision  
as to whether a given URI-R should be looked up in a specific archive.  
Specifically, we conduct experiments in which we use the content of the Aggregator Cache to train one classifier 
per archive covered by the Aggregator. The training is based on features extracted from the URI-Rs stored in the cache 
and uses the TimeMap information contained in the cache that indicates whether an archive holds Mementos for 
that URI-R or not. 
Once a classifier for an archive has been generated, it can provide a binary response 
to the question whether the archive should be polled for a given URI-R. 

If such an approach were successful 
in reducing the amount of distributed queries, it would be rather attractive from an operational perspective:
\begin{itemize}
\item Unlike previously explored approaches, it does not require the involvement of third-party data as it is fully based on available cached data.
\item As archive holdings, and hence the cached content evolves, classifiers can recurrently be retrained 
in off-line background processes without affecting overall Aggregator performance. In addition, 
since we generate the classifiers with fixed features types but dynamically 
selectable features and number of features per type, they can automatically adapt to a changing 
web archiving landscape.
\item It can be expected that the negligible overhead that would be incurred by realtime querying all classifiers 
(a fraction of milliseconds) would by far be offset by the benefits of not having to query all archives. 
\end{itemize}

The remainder of the paper is structured as follows: Section \ref{section:RelatedWork} provides an overview of prior 
work in this realm; Section \ref{section:BuildingArchiveSpecificClassifiers} describes how classifiers 
are generated and details the choice of training features and algorithms; Section \ref{section:Evaluation} 
provides an evaluation of the classifiers using a large dataset of URI-Rs that are distinct from those 
in the Aggregator cache; Section \ref{section:Conclusions} summarizes our findings.

\section{Related Work}
	\label{section:RelatedWork}

Optimizing Memento query routing has been explored in efforts that rely on archive profiling. 
In \cite{AlSum2014}, profiles were created based on top-level domain (TLD) that recorded URI-R and 
URI-M counts per TLD for twelve public web archives. The results show that it is possible to retrieve a complete TimeMap for 84\% 
of URI-R when using only the top 3 archives and in 91\% of the cases when using the top 6 archives. 
This simple approach can reduce the number of queries generated by a Memento aggregator significantly with some loss in coverage. 
In \cite{alam2015} extensive profiles were created based on URI keys, 
generated from URI-Rs using various templating approaches. 
Doing so, they can successfully identify about 78\% of URI-R to not be present 
in an archive by means of a template approach that requires storing only 1\% 
of what would be required to hold all URI-Rs of the archive. 
Both \cite{alam2015,AlSum2014} ideally require obtaining URI-R index files from archives.  
Profiles could also be generated by sampling archives for URI-Rs, although determining an appropriate 
sampling approach remains a research challenge in its own right.
While these research directions are interesting and promising, generating profiles is resource intensive, 
requires recurrent updates at unpredictable frequencies as archives evolve, 
and - in case of the index file approach - relies on the availability of third party data and, 
hence the willingness of those parties to share it.

Various efforts have used machine learning techniques to predict characteristics 
of a web page by merely considering its URI. 
The classification goals are wide ranging and include predicting a web page's topic  
\cite{Baykan:2009:PUT:1526709.1526880,Baykan:2011:CSF:1993053.1993057,Kan:2005:FWC:1099554.1099649}, 
 genre \cite{genreClassificationUrl}, pagerank \cite{Kan:2005:FWC:1099554.1099649},   
language \cite{Baykan:2013,Baykan:2008} or whether it has malicious content 
\cite{PhisingURLs,Whittaker,Ma:2009:ISU:1553374.1553462}.
Certain URI feature classes perform better for some goals than others. 
The lexical features of a URI were successfully used to detect phishing 
attacks \cite{PhisingURLs,Whittaker,Ma:2009:ISU:1553374.1553462}. 
TLD has been used for language detection \cite{Baykan:2013,Baykan:2008} but results show that, 
due to the heterogeneous nature of domains like com and org using TLD only is not sufficient. 
In \cite{Kan:2005:FWC:1099554.1099649}, several token segmentation techniques were used 
to determine web page topic.  
The resulting classifiers perform well on long URIs but less so on typical web site entry points. 
An approach that includes the use of tokens has also achieved high accuracy 
in identifying suspicious URIs  
\cite{PhisingURLs,Ma:2009:ISU:1553374.1553462}. For text classification, n-gram approaches are widely used  
and have also been applied for URI classification in combination with tokens  
for topic and genre classification \cite{genreClassificationUrl,Baykan:2009:PUT:1526709.1526880,Baykan:2011:CSF:1993053.1993057} 
as well as  for language detection \cite{Baykan:2013,Baykan:2008}. 
These efforts have achieved good results for their respective goals, and we build on their pioneering work. 
However, we apply their techniques to an entirely different task. 
As we embark on the research we are unsure whether it will be possible to characterize 
the respective archives by means of a limited set of features, 
especially since the holdings of many are highly heterogeneous, 
covering many languages, topics, and - in the case of on-demand archives - user interest. 

\section{Building Archive-Specific Classifiers}
	\label{section:BuildingArchiveSpecificClassifiers}

For the purpose of our experiment, we use a dump of the content of the LANL Aggregator Cache, created 
on September 8th 2015. It contains 1,219,999 distinct cached URI-Rs for which a total 
of 239,753,370 URI-Ms are known. Table \ref{table:URIdistribution} shows the number of cached URI-Rs 
for each archive as well as the number of cached URI-Rs for which an archive is the only one to hold Mementos. 
The Table shows that for 2 of the archives covered by the Aggregator (nara, si), 
the cache contains no URI-M at all. As a result, these archives are not included in  
the experiments as no training data is available for them (see Table \ref{table:ArchiveCoverage}). 
As can also be seen, for a large majority of URI-R, the Internet Archive (ia) holds Mementos. 
This observation is aligned both with prior findings and 
popular knowledge. As any sensible cross-archive lookup strategy would always include the Internet Archive, 
we decide not to train a classifier for this archive but 
rather to consistently perform a lookup, the equivalent to 
having a classifier that returns a positive, irrespective of the requested URI.
Overall, the Table clearly illustrates the value of 
looking beyond the Internet Archive when in need of a comprehensive overview of Memento holdings.

\begin{figure}[t]
\centering
	\includegraphics[width=0.40\textwidth]{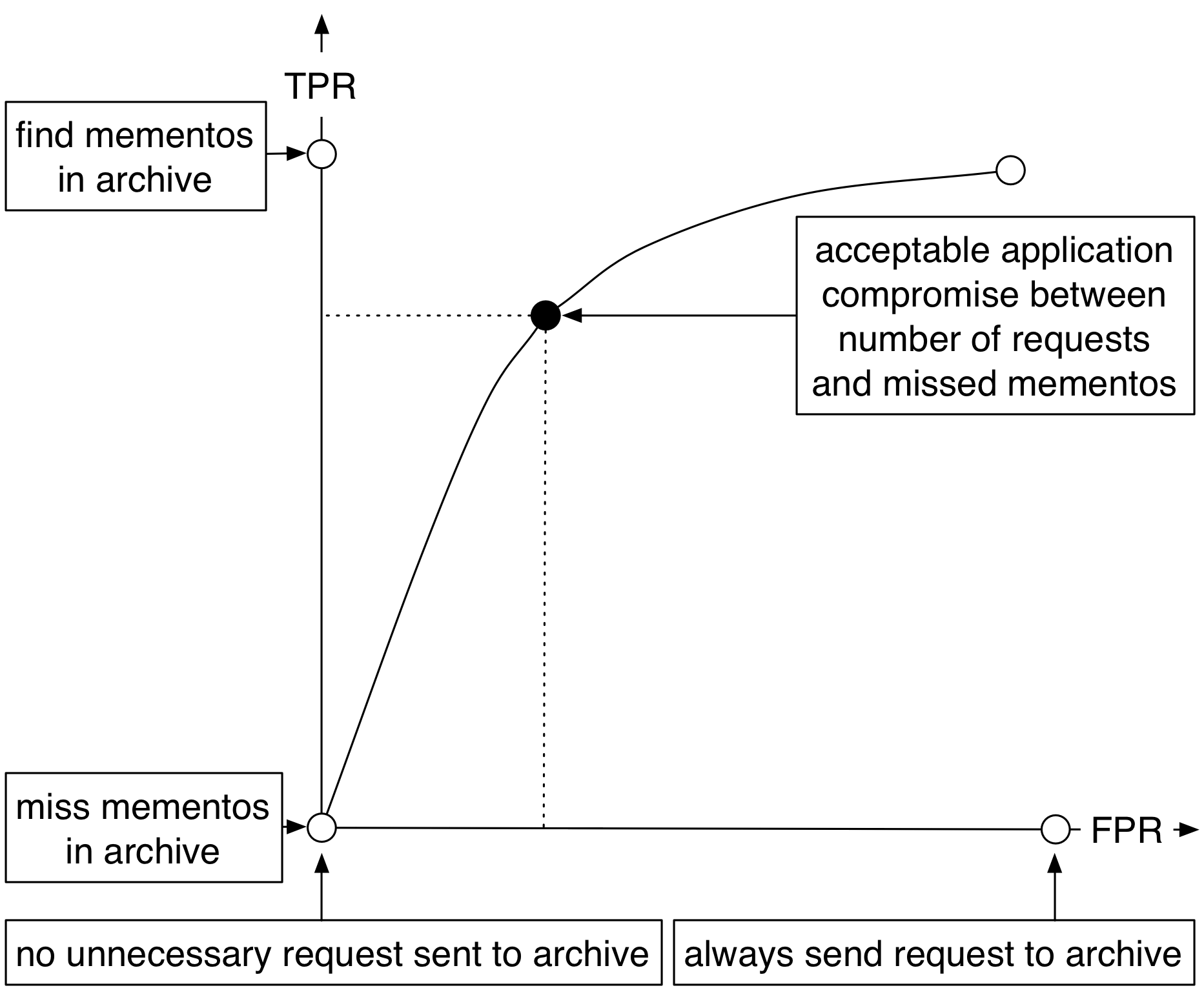}
	\caption{ROC curve for Memento requests to an archive}	
	\label{fig:ROCmemento}
\end{figure}

To visualize the performance of the archive-specific classifiers, we use Receiver Operating Characteristic 
(ROC) curves \cite{mitchell1997machine}. Figure \ref{fig:ROCmemento} illustrates the specific meaning of these curves for the case 
of routing Memento requests to an archive. In ROC curves, the x-axis represents the False 
Positive Rate (FPR) and the y-axis the True Positive Rate (TPR). 
When requesting a prediction from a trained classifier, 
a specific (TPR,FPR) pair is chosen on the curve that corresponds with the 
compromise that is most acceptable for a given application. 
Throughout the paper, we present ROC curves for two archives: 
the left hand plots are for archiveit that holds Mementos for a significant number of cached URI-R, 
and the right hand plots are for gcwa that holds Mementos for only a small number. 
To support a complete understanding, the ROC curves for all archives and all experiments 
are available\footnote{\url{http://mementoweb.org/demo/aggregator_learning/}}. 
To generate our classifiers, we use Apache Spark MLlib version 1.5.1 (scala)\footnote{\url{https://spark.apache.org/mllib/}} 
on a MacBook Pro, 2.7 Ghz i7, 16GB 1600Mhz DDR3 and use 10-fold cross-validation 
to train.

\begin{table}
    \caption[]{Distribution of the cached URI-R in the archives.}
    \centering
    \begin{tabular}{ccc}
        \toprule
            archive & \#URI-R stored & \#URI-R unique \\
        \midrule
            archive.is & 319,554 & 9,971 \\
            archiveit & 168,286 & 1,498 \\
            ba & 110,073 & 236 \\
            blarchive & 21,300 & 659 \\
            es & 4,170 & 50 \\
            gcwa & 1,001 & 10 \\
            hr & 1,245 & 0 \\
            ia & 920,934 & 390,604 \\
            is & 71,015 & 2,221 \\
            loc & 150,882 & 1,012 \\
            nara & 0 & 0 \\
            proni & 3,946 & 8 \\
            pt & 32,002 & 224 \\
            sg & 3,247 & 9 \\
            si & 0 & 0 \\
            swa & 895 & 8 \\
            uknationalarchives & 24,572 & 368 \\
            ukparliament & 14 & 1 \\
            webcite & 40,043 & 108 \\
        \bottomrule
        \label{table:URIdistribution}
    \end{tabular}
\end{table}

%

\subsection{Selecting Features}

Inspired by the aforementioned literature on using machine learning approaches for 
URI classification, we decide to use the following \emph{count} features: 
\begin{itemize}
  \item The character lengths of the complete URI-R and of its host, path, and query components.
  \item The count of special characters (\# / . ? - \_ ~ \% = : \$) in the aforementioned URI-R components.
\end{itemize}
Since the crawling policies of various archives differ, for example regarding depth of crawl, 
we expect these features to be relevant for our goal. 
Instead of using a Top Level Domain (TLD) feature as previous work did, we add the 
Public Suffix List domain\footnote{List at \url{https://publicsuffix.org/}} 
- \emph{PSL domains} - feature to our arsenal. It consists of a binary vector with 
one entry for each considered PSL domain. The extension from TLD to PSL is 
guided by the observation that most archives cover the same popular TLDs. 
In addition, we decide to also add \emph{n-gram} 
(n ranging from 3 to 7) and \emph{token} 
features extracted from URI-Rs as these have shown to be successful for determining the language of a web page. 
Since, especially, national archives may be more likely to archive web pages in certain languages, 
our intuition is that these features should add significant discriminatory power. 
Full word extraction (tokens) present a challenge in
our case as initial tests show that using dictionary lookups
is unsuccessful, a result of, for example, the use of trademarks and concatenated 
words in URI-Rs.
Hence, we decide on a simple approach that consists of generating tokens of length 2 to 10 by parsing a URI-R, 
removing common delimiters, and turning the resulting strings into lower case. 
Table \ref{table:ExampleFeatures} illustrates these features by means of an example URI-R. 

%
%

\begin{table*}
\caption{Example of features extracted for http://www.dailymail.co.uk/science-tech/index.html}
\label{table:ExampleFeatures}

\centering
\begin{tabular}{cc}
\toprule
Type & Features \\
\midrule
counts & len(url)=50, len(host)=19, count(., url)=4, count(., path)=1, ... \\
PSL domains & co.uk \\
3-grams on host & www, dai, ail, ily, lym, yma, ail \\
4-grams on path & scien, cien, ince, tech, inde, ndex, html \\
tokens on whole URI & www, dailymail, co, uk, science, tech, index\\
\bottomrule
\end{tabular}
\end{table*}

\begin{table}
        \caption{Observed and maximum features per type}
        \label{table:features}

    \centering
    \begin{tabular}{ccc}
        \toprule
        Features & Observed & Maximum \\
        \midrule
        counts & 36 & 36 \\
        PSL domains & 1,600 & 7,834 \\
        3-grams & 40,712 & 46,656 ($36^3$) \\
        4-grams & 345,988 & 1,679,616 ($36^4$) \\
        5-grams & 864,992 & 60,466,176 ($36^5$) \\
        2-10 tokens & 315,798 & - \\
        total & 1,569,126 & - \\
        \bottomrule
    \end{tabular}
\end{table}

\begin{table}
        \caption{Final features choice}
        \label{table:SelectedFeatures}

    \centering
    \begin{tabular}{ccc}
        \toprule
        Features Type & Number & Selection Metric \\
        \midrule
        counts & 36 & Take all \\
        PSL domains & 250 & Most Common \\
        3-Grams & 3,000 & Difference \\
        Tokens & 2,000 & Entropy \\
        \bottomrule
    \end{tabular}
\end{table}

\begin{figure*}[t]
\centering
\vspace{0.3cm}
\begin{subfigure}{.5\textwidth}
  \centering
  \caption*{archiveit}
  \includegraphics[width=0.8\linewidth]{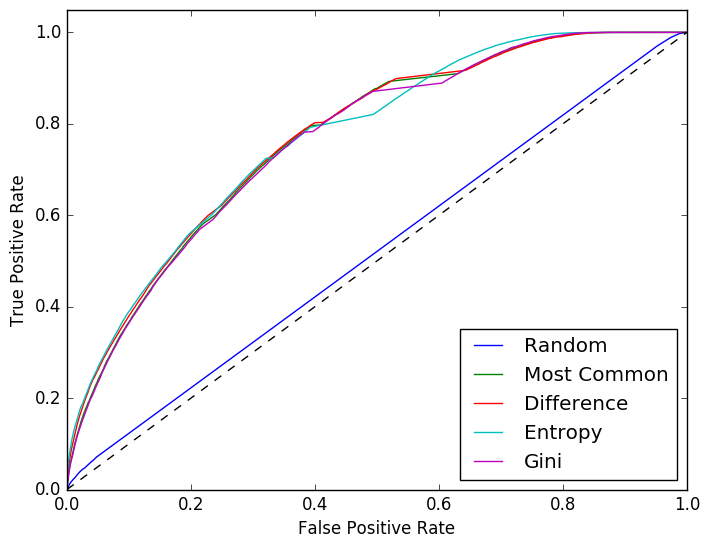}
\end{subfigure}%
\begin{subfigure}{.5\textwidth}
  \centering
  \caption*{gcwa}
  \includegraphics[width=0.8\linewidth]{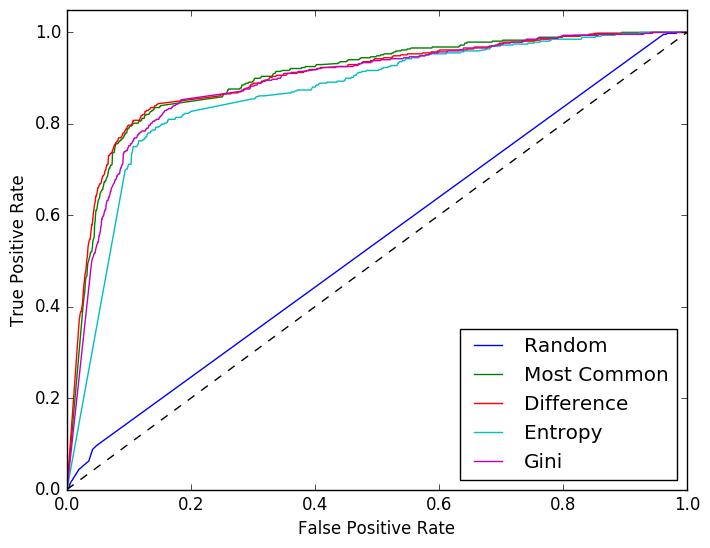}
\end{subfigure}
%
\centering
\begin{subfigure}{.5\textwidth}
  \centering
  \includegraphics[width=0.8\linewidth]{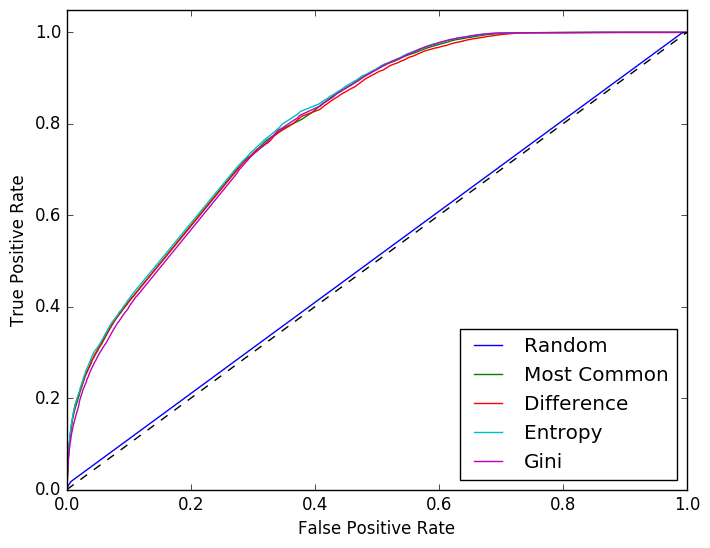}
\end{subfigure}%
\begin{subfigure}{.5\textwidth}
  \centering
  \includegraphics[width=0.8\linewidth]{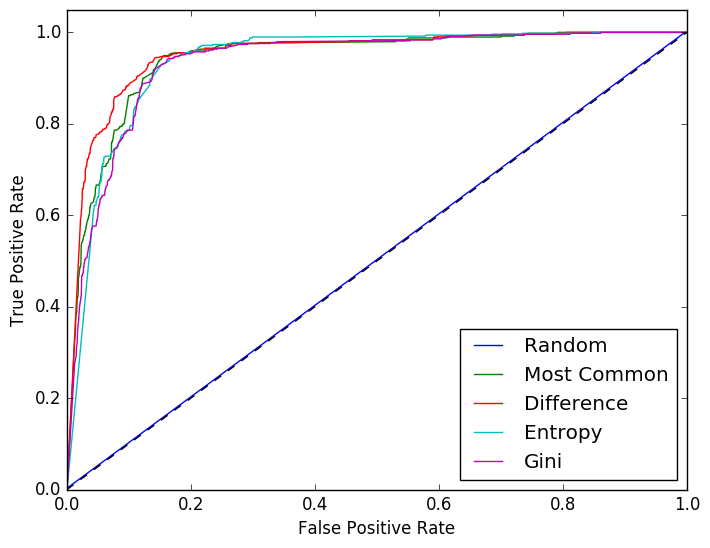}
\end{subfigure}
\caption{Comparison of feature selection strategies. Plots at the top: 
1,000 3-grams. Plots at the bottom: 1,000 tokens.}
\label{fig:MetricComparison}
\end{figure*}

Table \ref{table:features} shows the features discussed so far, and, for each, the  
number observed in the set of cached URI-Rs as well as the maximum, if any.
Since one of our goals is to incur minimal overhead by querying the archive-specific classifiers in realtime, 
it is not feasible to exploit all those features for classification. 
While we need not be concerned about the number of counts  
features, we do need to limit the number of PSL domains, n-gram, and token features. There are two aspects to the desired reduction: 
\begin{itemize}
  \item Selecting a method to rank features according to their discriminatory ability.
  \item Selecting the feature types to use, and, for each type, the maximum number of features. 
\end{itemize}

Regarding the selection of ranking methods for features, we explore 4 metrics: 
the Most Common metric simply ranks features according to their frequency over the whole
training set; Difference is the sum of the absolute differences between a feature's frequencies for URI-Rs stored by an 
archive and the overall frequency of that feature (as used by Most Common); Entropy \cite{hastie2009}; 
and Gini impurity \cite{breiman1984classification}. 
The latter two are widespread metrics for assessing the usefulness of a split when building decision trees.



%
%

To quantify how the choice of a metric impacts the resulting
prediction, we select 1,000 features for the n-grams and token categories according to the 4 aforementioned metrics, 
as well as 1,000 randomly selected features, to be used as a reference point. We then train binary classifiers
using the logistic regression algorithm.  
As the ROC curves of Figure \ref{fig:MetricComparison} illustrate, 
we find that the choice of metric does not significantly impact the resulting classifier.  
We observe the same lack of impact of the choice of metric for classifiers generated for all archives 
and find that it relates to the significant overlap in choice of features for each metric. For example, 
we find that when it comes to selecting 1,000 3-gram features for archiveit, 
the smallest overlap in features is between the Most Common and Entropy metrics, which still share 563 features. 
Nevertheless, we find small performance differences, leading us to proceed with the Most Common metric for PSL domains, 
Difference for n-grams, and Entropy for Tokens.

\begin{figure*}
\centering
\begin{subfigure}{.5\textwidth}
  \centering
  \caption*{archiveit}
  \includegraphics[width=0.8\linewidth]{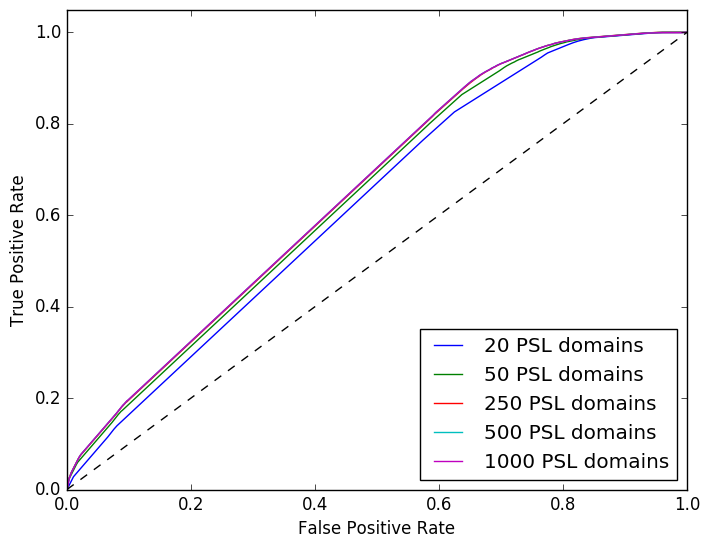}

\end{subfigure}%
\begin{subfigure}{.5\textwidth}
  \centering
  \caption*{gcwa}
  \includegraphics[width=0.8\linewidth]{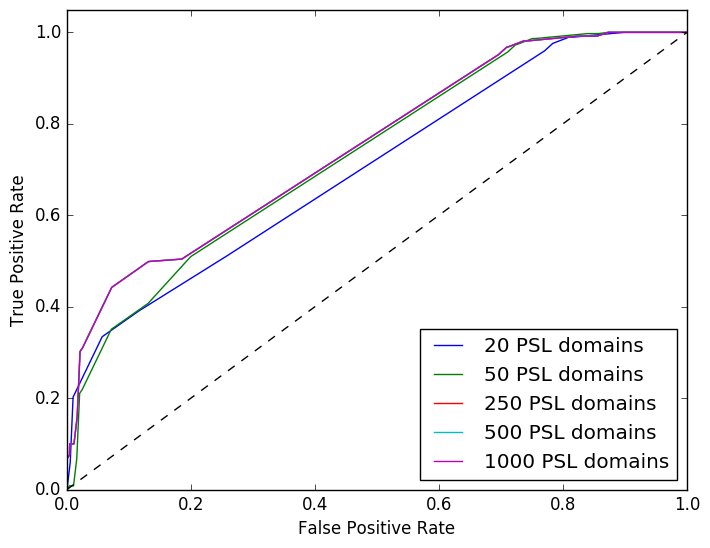}
\end{subfigure}

\centering
\begin{subfigure}{.5\textwidth}
  \centering
  \includegraphics[width=0.8\linewidth]{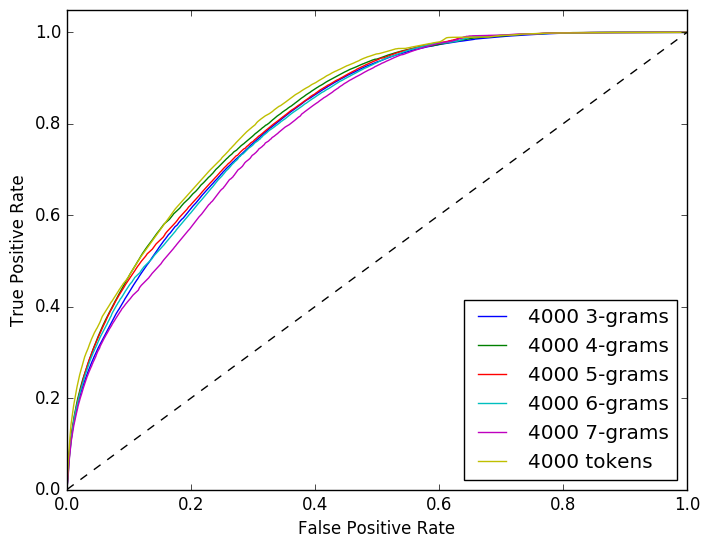}
\end{subfigure}%
\begin{subfigure}{.5\textwidth}
  \centering
  \includegraphics[width=0.8\linewidth]{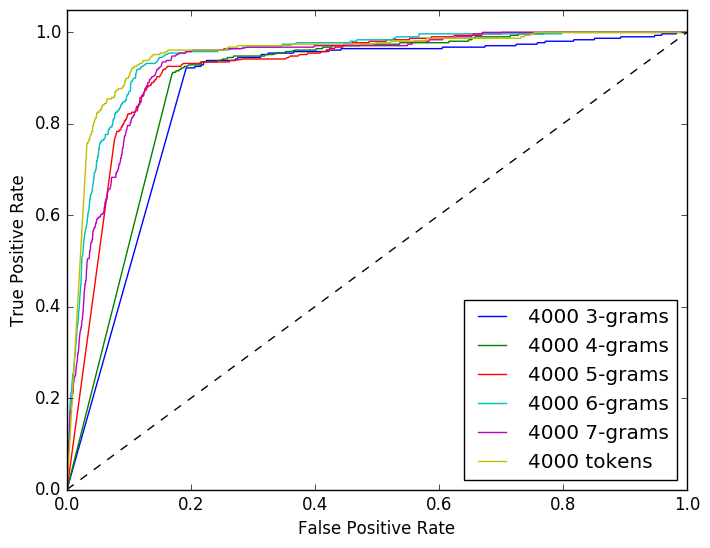}
\end{subfigure}

\centering
\begin{subfigure}{.5\textwidth}
  \centering
  \includegraphics[width=0.8\linewidth]{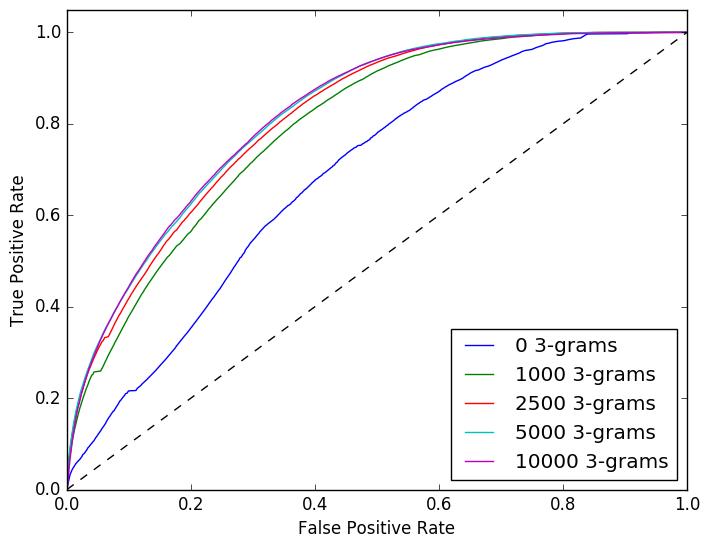}
\end{subfigure}%
\begin{subfigure}{.5\textwidth}
  \centering
  \includegraphics[width=0.8\linewidth]{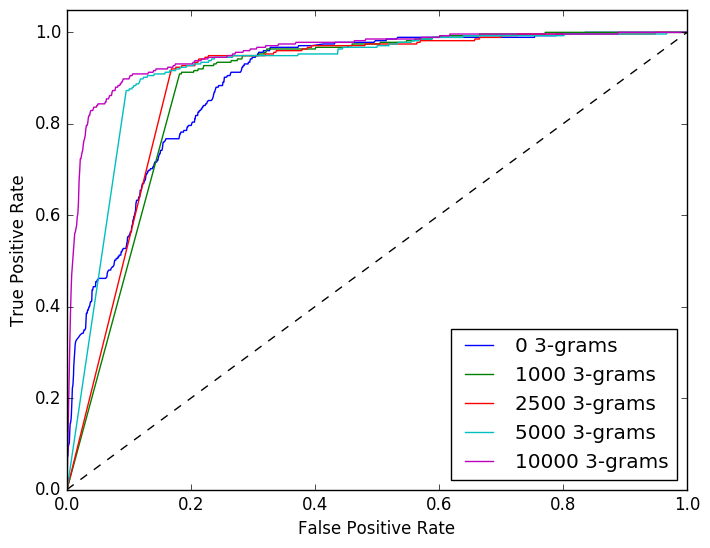}
\end{subfigure}

\centering
\begin{subfigure}{.5\textwidth}
  \centering
  \includegraphics[width=0.8\linewidth]{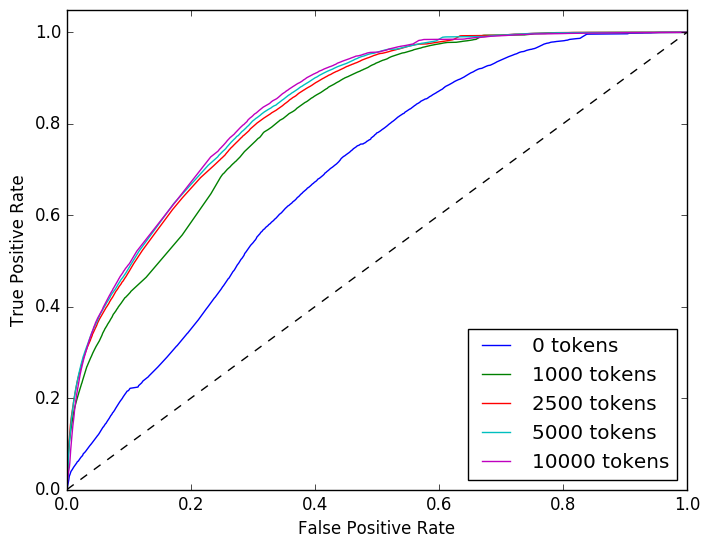}
\end{subfigure}%
\begin{subfigure}{.5\textwidth}
  \centering
  \includegraphics[width=0.8\linewidth]{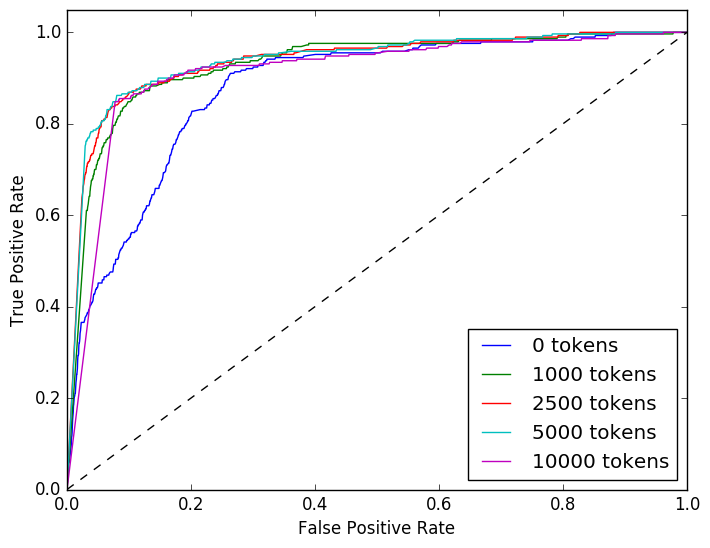}
\end{subfigure}

\caption{Comparison of number of features. Plots at the top: 
PSL domains. Plots second from top: n-grams and tokens. Plots third from top: 3-grams. Plots at the bottom: tokens.}
\label{fig:NumberFeatures}
\end{figure*}

Regarding the selection of types and numbers of features, we evaluate various scenarios for PSL domains, n-grams, and tokens. 
For each, we choose the respective metric resulting from the above described experiments, and, again,  
generate classifiers using the logistic regression algorithm to evaluate performance.
Regarding PSL domains, we explore the use of different numbers of features: 20, 50, 250, 500, and 1,000. 
Figure \ref{fig:NumberFeatures}, top, shows the resulting ROC curves. 
They illustrate a pattern that occurs for all archives, namely that performance does not increase significantly by using 
more than 250 PSL domain features, which is the number we select. 
We next focus on n-grams and tokens and proceed as follows: first, we compare the different types of features 
(e.g. 3-grams, 4-grams, tokens) to see whether some stand out; next, we determine the number of features per type. 
We find that 3-grams and 4-grams perform best (Figure \ref{fig:NumberFeatures}, second from top) and that a number of features between 2,500 
and 5,000 is desirable (Figure \ref{fig:NumberFeatures}, third from top for 3-grams, and bottom for tokens). 
After conducting more detailed assessments in the range 2,500-5,000, we decide to settle on 3,000 3-grams and 2,000 tokens. 
This is a somewhat arbitrary 
decision because adding more features further improves the predictions. 
However, the gains become too small to justify the additional computational cost. 
Table \ref{table:SelectedFeatures} summarizes the 
chosen features and respective numbers.

We conclude our exploration of features by assessing the performance of several feature combinations. 
As Figure \ref{fig:composeFeatures} shows, we find that performance can substantially be improved by using 3-grams and 
tokens in addition to the basic (counts and PSL domains) features. Of all the variations we try, 
it turns out that basic combined with 3-grams and tokens extracted from the whole URI-R perform best. \\\\

\subsection{Selecting Training Algorithms}

So far, we have used logistic regression only as the algorithm to train the classifiers. Here, 
we assess the performance of different algorithms using the features selected above. We are specifically 
interested in algorithms that result in classifiers that have a low computational load and small 
memory footprint at runtime. 
Hence, we exclude algorithms such as Nearest Neighbors that require the 
availability of the entire training set at runtime. The choice of the Spark framework, selected among others 
because its ability to deal with extensive datasets, further limits the choice of algorithms to 
Logistic Regression, Multinomial Bayes, Random Forest, and Support Vector Machine (SVM) with stochastic gradient descent. 

\begin{figure*}[t]
\centering

\begin{subfigure}{.5\textwidth}
  \centering
  \caption*{archiveit}
  \includegraphics[width=0.8\linewidth]{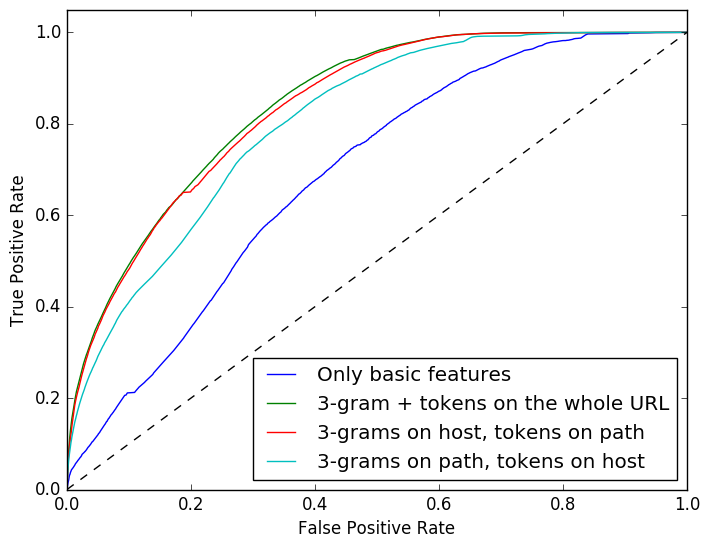}
\end{subfigure}%
\begin{subfigure}{.5\textwidth}
  \centering
  \caption*{gcwa}
  \includegraphics[width=0.8\linewidth]{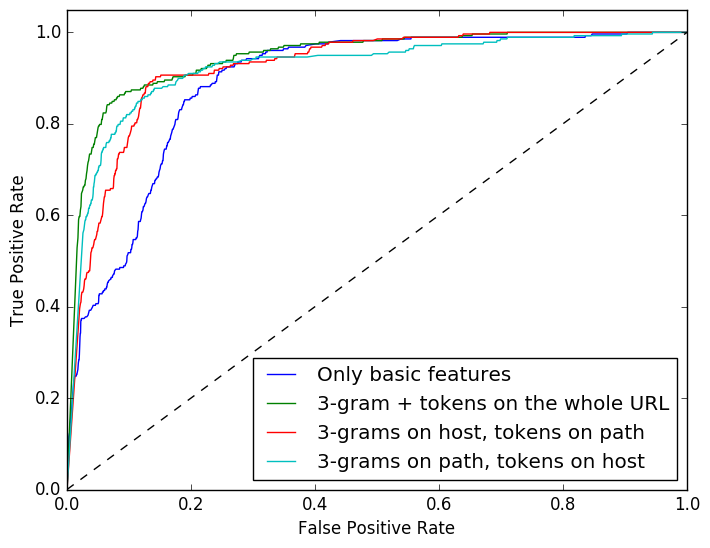}
\end{subfigure}
\caption{Combining basic (count, PSL domains), 3-grams, and token features}
\label{fig:composeFeatures}
%

\vspace{0.06cm}

\begin{subfigure}{.5\textwidth}
  \centering
  \caption*{archiveit}
  \includegraphics[width=0.8\linewidth]{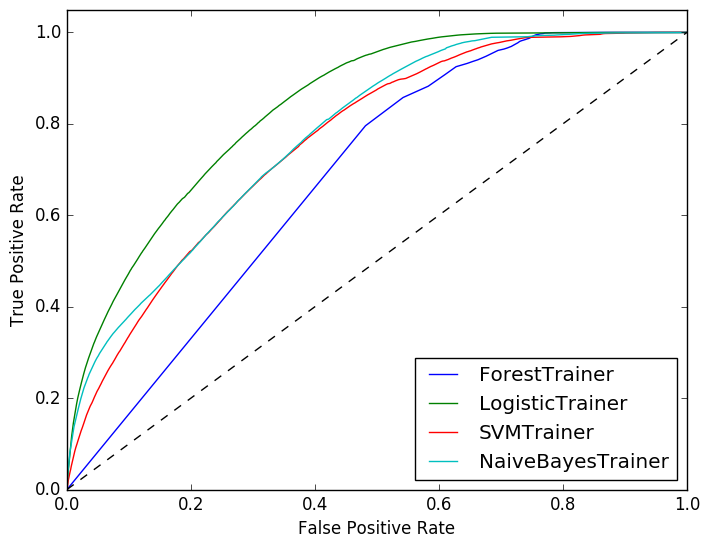}
\end{subfigure}%
\begin{subfigure}{.5\textwidth}
  \centering
  \caption*{gcwa}
  \includegraphics[width=0.8\linewidth]{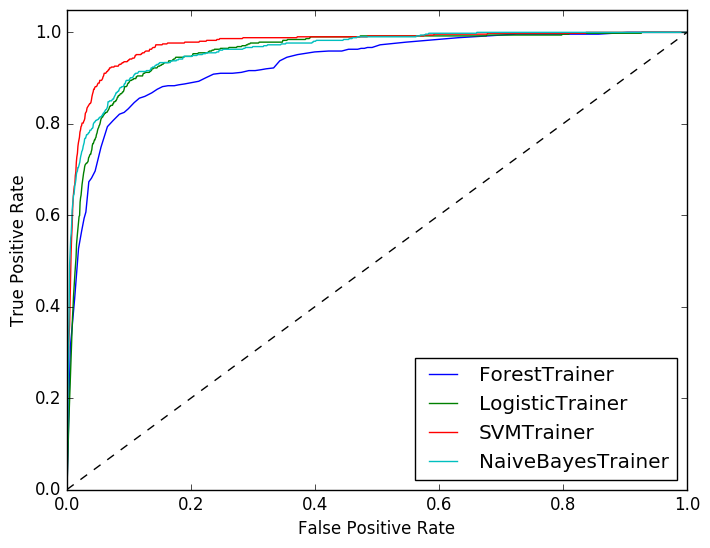}
\end{subfigure}
\caption{Comparison of training algorithms}
\label{fig:Algorithms}

\end{figure*}

Figure \ref{fig:Algorithms} shows the ROC curves whereas 
Table \ref{table:algorithmCompare} lists, per algorithm, the time  
required to train the classifier 
and to obtain 100K predictions. We find that Random Forest 
yields the worst results both regarding algorithm performance and prediction times; 
we therefore discard it. 
We find no clear winner among the remaining 3 algorithms. Their performance and runtime prediction times 
are very similar; the latter are negligible as anticipated. 
The learning times differ but are not a significant concern for our 
application because training can be done in offline processes. 
We proceed to train 3 classifiers per archive, one using each algorithm.
In preparation of evaluating their performance (see Section
\ref{section:Evaluation}), we need to determine the
thresholds under which the classifiers must perform in order to
achieve a targeted True Positive Rate (TPR).
We initially rely on a subset of the cached entries, distinct from the
training set, to determine these thresholds.
However, when evaluating the classifiers on third party URI samples,
we find that they are overly optimistic in the sense that they recommend
too few lookups. We assume this is related to the nature of pockets of
cached URI-Rs that share
the same baseURL, a result of users looking up batches of URIs for a
same domain. Hence,
we bring in an external dataset of 100,000 totally unrelated URI-Rs
extracted from log files
of the Internet Archive covering requests issued on January 27th 2012.
We use these URI-Rs to determine the threshold at which to query each
archive-specific classifier to achieve a required TPR,
and, for each archive, select the algorithm that yields the lowest
False Positive Rate (FPR).
We find that Logistic Regression performs best for 10 archives
(archiveit, ba, blarchive, es, loc, proni, pt, uknationalarchives,
ukparliament, webcite) and Multinomial Bayes for 6 (archive.is, gcwa,
hr, is, sg, swa).
The inclusion of this external data is somewhat of a setback since we had hoped to fully rely on cached data only. 
Nevertheless, we note that this dataset can be the same 
for recurrent classifier training as long as associated Memento information would recurrently be updated. 
Such information can be gathered using TimeGate requests, which are cheaper than TimeMap requests. 
Also, this information has shown to evolve slowly 
over time\cite{jcdl13:timemaps}, making polling these URI-Rs for each recurrent classifier training unnecessary, 
although this finding would need to be reconfirmed.

\begin{table}[t]
    \centering
    \begin{tabular}{ccc}
        \toprule
        Algorithm & Learning & 100K Predictions \\
        	&	Time (s) &  Time (s) \\
        \midrule
        Logistic Regression & 18.47 & 0.609 \\
        Multinomial Bayes & 5.14 & 0.487 \\
        Random Forest & 76.13 & 11.45 \\
        SVM & 261.94 & 0.48 \\
        \bottomrule
    \end{tabular}
\caption{Learning time averages over all archives and 3 runs}
\label{table:algorithmCompare}
\end{table}

\section{Evaluation}
	\label{section:Evaluation}

Having trained the classifiers,  
we proceed to evaluate their 
performance using an unrelated datasets consisting of URI-Rs extracted from logs of oldweb.today covering 200,000 randomly selected requests 
issued in the week of December 14th 2015. We remove URI-Rs that are syntactically invalid, 
duplicate, already covered by our cache, or blocked by our adult-content filters. The resulting set has 
187,449 URI-Rs. Since these originate from requests issued to a service 
that operates across archives, and 
are not covered by our cache, they are representative of the URI-Rs for which the Aggregator 
infrastructure would need to send distributed requests to archives in order to assemble an aggregate response. 

To evaluate the performance of the classifier-based approach to sending requests, for each URI-R, we:
\begin{itemize}
  \item Issue a TimeMap request to determine which archives hold associated Mementos.
  \item Query each archive-specific classifier to determine whether it advises a lookup in the archive or not. 
  We query the respective classifiers at several FPR levels: 0.9, 0.8, 0.7, 0.6, and 0.5. 
  \item Assess recall, computational cost, and response times using the obtained data. 
\end{itemize}

We use the common definition of recall ($\frac{TP}{TP+FN}$), with 
TP being True Positive, and FN False Negative.
To assess computational cost, we use the sum of the time it takes to poll each archive recommended by the classifiers 
for a given URI-R as this relates to the load on the Aggregator infrastructure and on the archives.
To assess response times experienced by a user of Aggregator services, 
we take the maximum response time over the archives polled for a given URI-R. 
In order to avoid issuing 
hundreds of thousands of requests to archives, we simulate the response time for a given URI-R per archive.
To do so, we collect 1,000 response times per archive. Table \ref{table:responseTime} 
shows the range of observed response times, listing minimum, average, and maximum. 
Then, for our computations, we randomly select with replacement - per archive and per URI-R - a response time from the 1,000 observed ones. 
That selected response time is used for classifiers operated at the different levels of TPR. 
Table \ref{table:recapTable} shows the results, based on the 187,449 URI-Rs from oldweb.today,  
distinguishing between using all archives or only Memento compliant ones. 
Note that the FPR value of 1.0 corresponds to not using classifiers but rather the brute force approach. 
The results indicate that the currently used heuristic to query all Memento-compliant archives yields the best recall  
(disregarding the brute force approach on all archives), yet that computational cost and response time can be reduced 
by using classifiers without significantly decreasing recall. Viable strategies exist both using all archives 
or compliant ones only, but the latter consistently perform better regarding recall and response time at equivalent request numbers. 
The result at TPR 0.6, using Memento compliant archives only, looks extremely attractive: 
compared to a brute force approach on all archives, 
classifiers can reduce the average number of requests by 77\% (from 17 to 3.985), 
and the overall response time by 42\% (from 3.712 to 2.16 seconds) while maintaining a recall of 0.847. At this TPR level, 
significant optimizations can be achieved while maintaining acceptable recall, even when 
compared to brute force on Memento compliant archives only.
When operating at FPR level 0.45, we reach an average number of 
2.994 requests per URI-R and find that 
complete TimeMaps are collected for 83.4\% of URI-R. This result fully aligns 
with \cite{AlSum2014}, which found that it is possible to retrieve a complete 
TimeMap for 84\% of URI-R when using only the top 3 archives.  
But, in contrast to \cite{AlSum2014}, our approach 
only marginally relies on third party data, 
and can actually be brought into production.

\begin{table}[t]
    \caption[]{Response time [ms] statistics}
    \centering
    \begin{tabular}{cccc}
        \toprule
            Archive & Min & Average & Max \\
        \midrule
            archive.is & 35 & 434 & 2,770 \\
            archiveit & 140 & 342 & 9,585 \\
            ba & 226 & 1,740 & 60,372 \\
            blarchive & 338 & 562 & 59,087 \\
            es & 438 & 464 & 1,387 \\
            gcwa & 219 & 464 & 2,516 \\
            hr & 407 & 428 & 2,817 \\
            ia & 71 & 1,485 & 24,967 \\
            is & 402 & 838 & 3,215 \\
            loc & 191 & 381 & 3,804 \\
            proni & 181 & 234 & 5,793 \\
            pt & 57 & 821 & 9,328 \\
            sg & 443 & 836 & 9,035 \\
            swa & 2 & 3 & 352 \\
            uknationalarchives & 190 & 308 & 6,320 \\
            ukparliament & 186 & 312 & 32,278 \\
            webcite & 495 & 1217 & 60,050 \\
        \bottomrule
        \label{table:responseTime}
    \end{tabular}
\end{table}

\begin{table}[t]
	\small
    \caption{Average (\#requests, recall, sum(T), max(T)) per URI-R on oldweb.today sample, with T the response time [s]}
    \centering
    \begin{tabular}{lcc}
        \toprule
            TPR & All archives & Memento compliant archives \\
        \midrule
            1.0 & (17.00, 1.000, 10.90, 3.712) & (11.00, 0.971, 6.640, 3.084) \\
            0.9 & (9.134, 0.955, 6.533, 2.983) & (6.447, 0.929, 4.506, 2.558) \\
            0.8 & (7.429, 0.924, 5.562, 2.760) & (5.384, 0.900, 3.995, 2.409) \\
            0.7 & (6.213, 0.896, 4.792, 2.534) & (4.619, 0.874, 3.597, 2.283) \\
            0.6 & (5.220, 0.867, 4.233, 2.418) & (3.958, 0.847, 3.229, 2.160) \\
            0.5 & (4.303, 0.835, 3.614, 2.226) & (3.349, 0.818, 2.867, 2.041) \\
        \bottomrule
        \label{table:recapTable}
    \end{tabular}
\end{table}

We zoom in on the 0.6 TPR level. 
For that level, Table \ref{table:evaluationSetup} shows, per archive, 
the true positives (TP), false negatives (FN), 
true negatives (TN), and false positives (FP). 
Note that TP+FN for an archive is equal 
to the number of URI-R of the sample for which the archive holds Mementos. 
Also, TP+FP is the number of queries sent to an archive.
For sg and ukparliament, only FP is listed as neither archive has Mementos 
for URI-Rs in the oldweb.today dataset. Since a request is always sent to ia, no FN are listed. 
Note, for ia, the significant number of URI-R for which it has no Mementos. 
Table \ref{table:NumberRequests} compares the number of requests sent according to 
various strategies. 
We see that, when including all archives, 
the classifiers at TPR level 0.6 recommend sending a total of 916,881 requests: 
171,862 TP and 745,019 FP. The high FP count relates to our desire to achieve low FN and hence 
miss few Mementos; FN stands at 26,304. The total number of requests 
would have been 3,186,633 for the brute force approach (TPR 1.0) on all archives.  
In this case, the classifiers achieve a 71\% reduction. 
When only Memento compliant archives are considered a reduction of 67\% is achieved. 

\begin{table}
\caption{Performance on oldweb.today dataset at TPR 0.6}
\label{table:evaluationSetup}
\centering
\begin{tabular}{lccccccc}
\toprule
        Archive  &  TP  &  FN  &  TN  &  FP \\
      \midrule      
archive.is & 14 & 62 & 185,541 & 2,518 \\
archiveit & 7,694 & 4,927 & 124,580 & 50,934 \\
ba & 19,888 & 9,593 & 95,988 & 62,666 \\
blarchive & 1,665 & 582 & 131,985 & 53,903 \\
es & 670 & 284 & 135,254 & 51,927 \\
gcwa & 210 & 113 & 149,161 & 38,651 \\
hr & 0 & 3 & 176,272 & 11,860 \\
ia & 122,787 & 0 & 65,348 & 0 \\
is & 5,362 & 2,381 & 94,760 & 85,632 \\
loc & 6,625 & 4,769 & 111,518 & 65,223 \\
proni & 489 & 336 & 140,201 & 47,109 \\
pt & 2,289 & 909 & 119,650 & 65,287 \\
sg & 0 & 0 & 188,135 & 0 \\
swa & 2,320 & 1,239 & 93,093 & 91,483 \\
uknationalarchives & 1,185 & 531 & 134,500 & 51,919 \\
ukparliament & 0 & 0 & 188,135 & 0 \\
webcite & 664 & 575 & 120,989 & 65,907 \\
Total & 171,862 & 26,304 & 2,255,110 & 745,019 \\        
    \bottomrule
\end{tabular}
\end{table}

\begin{table}
\caption{Number of requests using different strategies}
\label{table:NumberRequests}
\centering
\begin{tabular}{lccccccc}
\toprule
        TPR & \#Requests &  \#Requests \\
        & all archives & Memento compliant \\
      \midrule      
1.0 & 3,186,633 & 2,061,393 \\
0.6 & 916,881 & 676,884  \\      
    \bottomrule
\end{tabular}
\end{table}

\begin{figure*}
\centering
\begin{subfigure}{.5\textwidth}
  \centering
  \includegraphics[width=1\linewidth]{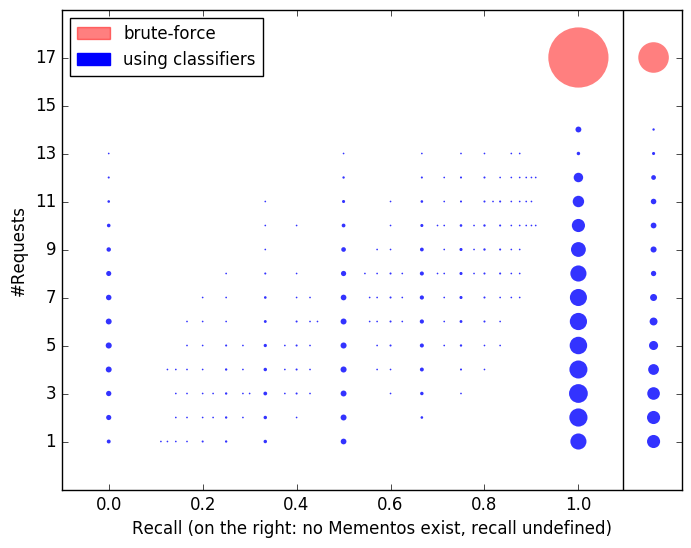}
\end{subfigure}%
\begin{subfigure}{.5\textwidth}
  \centering
  \includegraphics[width=1\linewidth]{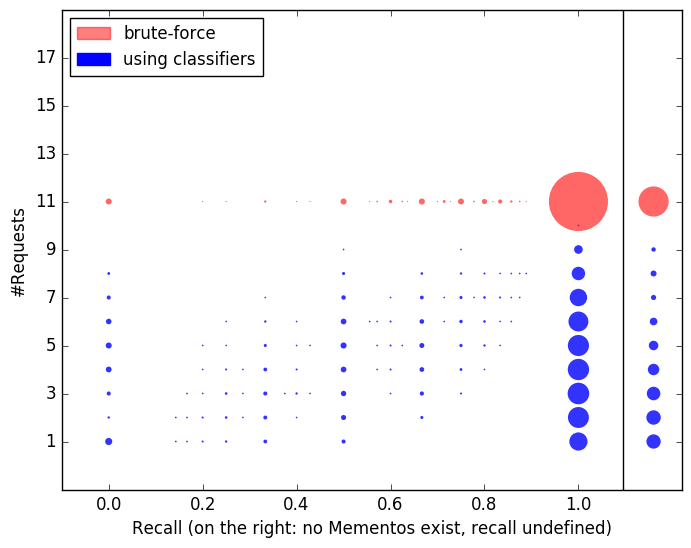}
\end{subfigure}
\caption{Recall per URI-R versus number of requests sent using all archives (left) and compliant archives (right). Dot size proportional to number of URI-Rs.}
\label{fig:ScatterPlots}
\end{figure*}

Figure \ref{fig:ScatterPlots} details the relation between recall 
and the number of requests sent, again for 
classifiers operating at 0.6 TPR. 
The left hand plot considers a situation in which all archives are involved, the right hand one pertains to Memento compliant ones only. 
In each case, brute force requests are depicted in red and requests based on the advise of classifiers in blue. 
Each plot covers all URI-Rs of the oldweb.today dataset and the size of the respective dots is proportional to the number of 
URI-R for a given (recall,requests) combination. The dots at the very right hand side of each plot pertain 
to URI-R for which no Mementos exist in any archive, and, hence, for which recall is undefined.
We see a very significant number of URI-R for which classifiers reach the maximum recall by sending between 1 and 9 requests 
but also some URI-R for which Mementos are missed even when sending up to 13 
requests. 

\section{Conclusions}
	\label{section:Conclusions}
	
We explored the use of binary classifiers to guide the routing of Memento requests for Memento Aggregators. 
To train the classifiers, we solely relied on 
data that is recurrently gathered by the LANL Aggregator as part of its daily operation. We used features that have been 
shown to perform well for other URI-based classifier tasks and 
determined a combination of number and types of features that worked well for the novel challenge of routing Memento queries.
We also trained archive-specific classifiers using various training algorithms on the basis of the same data. 
However, in order to optimally operate the classifiers, 
we had to bring in a third party set of URI-Rs to compensate for bias in the Aggregator Cache. 
Our evaluation of this approach, performed on the basis of an unrelated set of URI-Rs from oldweb.today, 
shows that classifiers can significantly reduce the number of requests sent to archives, 
and hence reduce the load on both the Aggregator 
and the archives. It can also reduce overall response times. 
These reductions can be achieved without significantly compromising recall. 
Improvements over the reported work are definitely possible. 
We must ensure that the cache contains URI-Rs with associated Mementos in all archives as the lack of training data 
led us to exclude two from our experiments. 
More advanced machine learning techniques can be explored that may yield even better results. 
But, overall, the results are so compelling that we already devised a workflow based on Spark that can recurrently train 
archive-specific classifiers on the basis of cached data. The training of classifiers is set up 
such that it can dynamically adapt with regard to specific features and number of features, as 
the archives evolve. We plan to bring this capability in production to guide the LANL Aggregator and will 
also expose a public API to support Memento clients in determining which archives to poll for a given URI-R. 

\section{Acknowledgments}
This work is supported in part by the International Internet
Preservation Consortium (IIPC). Ilya Kreymer provided the oldweb.today dataset. 
Shawn Jones, Martin Klein, and Harihar Shankar provided comments to a draft paper.
%

%
%
\bibliographystyle{abbrv}
\bibliography{paper-main-arxiv}  

\begin{thebibliography}{10}

\bibitem{Baykan:2013}
{A Comprehensive Study of Techniques for URL-based Web Page Language
  Classification}.
\newblock {\em ACM Trans. Web}, 7(1), 2013.

\bibitem{genreClassificationUrl}
M.~Abramson and D.~W. Aha.
\newblock {What is in a URL? Genre Classification from URLs}.
\newblock Technical Report AAAI Technical Report WS-12-09, Naval Research
  Laboratory, Washington, DC 20375, 2012.

\bibitem{alam2015}
S.~Alam, M.~L. Nelson, H.~Van~de Sompel, L.~Balakireva, H.~Shankar, and
  D.~Rosenthal.
\newblock {Web Archive Profiling Through CDX Summarization}.
\newblock In S.~Kapidakis, C.~Mazurek, and M.~Werla, editors, {\em Research and
  Advanced Technology for Digital Libraries}, volume 9316 of {\em Lecture Notes
  in Computer Science}, pages 3--14. Springer International Publishing, 2015.

\bibitem{AlSum2014}
A.~AlSum, M.~C. Weigle, M.~L. Nelson, and H.~Van~de Sompel.
\newblock {Profiling Web Archive Coverage for Top-Level Domain and Content
  Language}.
\newblock {\em International Journal on Digital Libraries}, 14(3):149--166,
  2014.

\bibitem{PhisingURLs}
R.~B. Basnet, A.~H. Sung, and Q.~Liu.
\newblock {Learning to detect phishing URLs}.
\newblock {\em {International Journal of Research in Engineering and
  Technology}}, 3(6):11--24, 2014.

\bibitem{Baykan:2009:PUT:1526709.1526880}
E.~Baykan, M.~Henzinger, L.~Marian, and I.~Weber.
\newblock {Purely URL-based Topic Classification}.
\newblock In {\em {Proceedings of the 18th International Conference on World
  Wide Web}}, WWW '09, pages 1109--1110, New York, NY, USA, 2009. ACM.

\bibitem{Baykan:2011:CSF:1993053.1993057}
E.~Baykan, M.~Henzinger, L.~Marian, and I.~Weber.
\newblock {A Comprehensive Study of Features and Algorithms for URL-Based Topic
  Classification}.
\newblock {\em ACM Trans. Web}, 5(3):15:1--15:29, July 2011.

\bibitem{Baykan:2008}
H.~M. Baykan~E. and W.~I.
\newblock {Web Page Language Identification Based on URLs}.
\newblock In {\em {Proceedings of the VLDBEndowment}}, volume 1(1), pages
  176--187, 2008.

\bibitem{breiman1984classification}
L.~Breiman, J.~Friedman, C.~J. Stone, and R.~A. Olshen.
\newblock {\em {Classification and Regression Trees}}.
\newblock CRC press, 1984.

\bibitem{jcdl13:timemaps}
J.~F. Brunelle and M.~L. Nelson.
\newblock An evaluation of caching policies for memento timemaps.
\newblock In {\em JCDL '13: Proceedings of the 13th ACM/IEEE-CS Joint
  Conference on Digital Libraries}, pages 267--276, 2013.

\bibitem{Whittaker}
M.~N. C.~Whittaker, B.~Ryner.
\newblock {Large-scale automatic classification of phishing pages}.
\newblock In {\em Proc. 17th Annual Network and Distributed System Security
  Symposium}, 2010.

\bibitem{hastie2009}
T.~Hastie, R.~Tibshirani, J.~Friedman, and J.~Franklin.
\newblock {\em {The Elements of Statistical Learning: Data Mining, Inference,
  and Prediction}}.
\newblock Springer, 2009.

\bibitem{Kan:2005:FWC:1099554.1099649}
M.-Y. Kan and H.~O.~N. Thi.
\newblock {Fast Webpage Classification Using URL Features}.
\newblock In {\em {Proceedings of the 14th ACM International Conference on
  Information and Knowledge Management}}, CIKM '05, pages 325--326, New York,
  NY, USA, 2005. ACM.

\bibitem{Ma:2009:ISU:1553374.1553462}
J.~Ma, L.~K. Saul, S.~Savage, and G.~M. Voelker.
\newblock {Identifying Suspicious URLs: An Application of Large-scale Online
  Learning}.
\newblock In {\em Proceedings of the 26th Annual International Conference on
  Machine Learning}, ICML '09, pages 681--688, New York, NY, USA, 2009. ACM.

\bibitem{mitchell1997machine}
T.~M. Mitchell.
\newblock {\em Machine Learning}.
\newblock McGraw-Hill Boston, MA:, 1997.

\bibitem{rfc7089}
H.~Van~de Sompel, M.~L. Nelson, and R.~Sanderson.
\newblock {HTTP Framework for Time-Based Access to Resource States -- Memento},
  December Internet RFC 7089, December 2013.

\bibitem{mementofirst}
H.~Van~de Sompel, M.~L. Nelson, R.~Sanderson, L.~Balakierva, S.~Ainsworth, and
  H.~Shankar.
\newblock {Memento: Time Travel for the Web}.
\newblock Technical Report arXiv:0911.1112, 2009.

\end{thebibliography}
\end{document}